\documentclass[%
 reprint,
 superscriptaddress,
preprintnumbers,
nofootinbib,
 amsmath,amssymb,
 aps,
]{revtex4-1}

\usepackage{amsmath}
\usepackage{amssymb}
\usepackage{amsthm}
\usepackage{MnSymbol}
\usepackage{nicefrac}
\usepackage{amsfonts}
\usepackage{dsfont}

\usepackage[markup=nocolor]{changes}
\usepackage{natbib}

\usepackage{graphicx}
\graphicspath{{Figures/}}  

\usepackage[normalem]{ulem}

\usepackage{dcolumn}
\usepackage{bm}
\usepackage{color}
\usepackage{slashed}
\usepackage{empheq}
\newcommand{\bea}{\begin{eqnarray}}
\newcommand{\eea}{\end{eqnarray}}
\newcommand{\be}{\begin{equation}}
\newcommand{\ee}{\end{equation}}

\begin{document}

\title{Quantum-gravity predictions for the fine-structure constant}
 
 \author{Astrid Eichhorn}
   \email{a.eichhorn@thphys.uni-heidelberg.de}
\affiliation{Institut f\"ur Theoretische
  Physik, Universit\"at Heidelberg, Philosophenweg 16, 69120
  Heidelberg, Germany}
\author{Aaron Held}
\email{a.held@thphys.uni-heidelberg.de}
\affiliation{Institut f\"ur Theoretische
  Physik, Universit\"at Heidelberg, Philosophenweg 16, 69120
  Heidelberg, Germany}
  \author{Christof Wetterich}
\email{c.wetterich@thphys.uni-heidelberg.de}
\affiliation{Institut f\"ur Theoretische
  Physik, Universit\"at Heidelberg, Philosophenweg 16, 69120
  Heidelberg, Germany}

\begin{abstract}
Asymptotically safe quantum fluctuations of gravity can 
uniquely determine the value of the gauge coupling for a large class of grand unified models. In turn, this makes 
the electromagnetic fine-structure constant calculable. The balance of gravity and matter fluctuations results in a fixed point for the running of the gauge coupling. It is approached as the momentum scale is lowered in the transplanckian regime, leading to a uniquely predicted value of the gauge coupling at the Planck scale.
The precise value of the predicted fine-structure constant depends on the matter content of the grand unified model. It is proportional to the gravitational fluctuation effects for which computational uncertainties remain to be settled.
\end{abstract}

\pacs{Valid PACS appear here}

\maketitle
Numerous speculations have attempted to explain the value of the fine-structure constant for decades. Within the standard model the renormalization flow of the U(1) gauge coupling has a Landau pole, indicating incompleteness at very high scales.
This problem is absent if the electroweak interactions are embedded in an asymptotically free grand unified theory (GUT). In an asymptotically free setting the value of the gauge coupling is, however, a free parameter and cannot be predicted. On the other hand, for a large number of matter fields, $\mathcal{N}>\mathcal{N}_c$, asymptotic freedom is lost and the Landau pole is back for any nonzero value of the gauge coupling $\alpha$ at the scale of spontaneous GUT-symmetry breaking $M_{\rm GUT}$. We argue that additional fluctuations of gravitational degrees of freedom in the transplanckian range of momenta beyond the Planck mass $M$ change the situation dramatically. For $\mathcal{N}>\mathcal{N}_c$, the flow of $\alpha$ is driven towards a fixed point $\alpha_{\ast}$, such that $\alpha(M)= \alpha_{\ast}$ becomes predictable. The fixed point defines ``gravi-gauge theories" as non-perturbatively renormalizable interacting quantum field theories (QFT) for gauge theories and quantum gravity.
In addition, an infrared (IR) unstable fixed point at $\alpha=0$ is present for all $\mathcal{N}> \mathcal{N}_c$.

Grand Unified Theories are an attractive scenario for beyond-standard-model physics, as they explain exact charge neutrality of atoms by unifying the leptons and quarks into combined representations \cite{Georgi:1974sy, Pati:1974yy, Fritzsch:1974nn}.  The gauge group SO(10) adds a right-handed neutrino to each generation and provides a simple explanation for small neutrino masses. Other parts of the necessary beyond-standard-model physics, such as the generation of baryon asymmetry, can  find an explanation within GUTs. The observed value of the Higgs boson mass and absence of other new physics at the LHC may suggest that the standard model could be valid
to considerably higher scales \cite{Shaposhnikov:2009pv,Bezrukov:2012sa}. Approximate unification of gauge couplings for energies around $10^{16}\, \rm GeV$ adds to the credibility of GUT scenarios. It is this unified gauge coupling whose value at the Planck scale $M$ we aim to predict. For energies below $M$, the perturbative renormalization flow of gauge theories without gravity relates $\alpha(M)$ to the observed low- energy couplings such as the electromagnetic fine-structure constant.
\\
In the asymptotic-safety scenario \cite{Weinberg:1980gg}, the high-energy behavior of running couplings is dominated by an interacting Renormalization Group (RG) fixed point,
 at which the scale-dependence vanishes. In contrast to asymptotic freedom, asymptotic safety can be realized for perturbatively renormalizable \cite{Litim:2014uca} as well as perturbatively non-renormalizable models  \cite{Gies:2003dp}. The asymptotic-safety scenario provides a framework in which a GUT-extension of 
the standard model could be combined with a QFT of the metric into an ultraviolet (UV) complete, predictive model of quantum gravity and matter \cite{Shaposhnikov:2009pv,Eichhorn:2017ylw,Eichhorn:2017lry}. Based on the groundbreaking work by Reuter \cite{Reuter:1996cp},  functional Renormalization Group \cite{Wetterich:1992yh} studies have accumulated compelling evidence for the viability of asymptotic safety in gravity \cite{Litim:2003vp, Benedetti:2009rx, Falls:2013bv, Becker:2014qya, Gies:2016con, Denz:2016qks}. The exploration of gravity-matter models has produced tantalizing hints that an ultraviolet complete model of matter and gravity could exist  \cite{Daum:2009dn,Narain:2009fy, Folkerts:2011jz,Harst:2011zx,Eichhorn:2012va, Oda:2015sma, Dona:2013qba,Dona:2015tnf, Eichhorn:2016vvy, Meibohm:2015twa,Eichhorn:2016esv,Christiansen:2017gtg, Biemans:2017zca,Hamada:2017rvn, Eichhorn:2017eht, Eichhorn:2017ylw,Eichhorn:2017lry,Christiansen:2017cxa} and  could explain some of the free parameters of the standard model \cite{Shaposhnikov:2009pv,Harst:2011zx,Wetterich:2016uxm,Eichhorn:2017ylw,Eichhorn:2017lry,Eichhorn:2017egq,Held:20175}. 

\subsection{Mechanism}
For GUTs without gravity the
 1-loop beta function for the running gauge coupling $\alpha(k) = \frac{g(k)^2}{4\pi}$ at renormalization scales $k$ beyond $M_\text{GUT}$ is given by
\be
k\partial_k\, \alpha=\beta_{\alpha}=\left(\mathcal{N}-
\mathcal{N}_c\right)\frac{\alpha^2}{4\pi}\;.
\ee
 For SO(10) the critical value $\mathcal{N}_c$ equals $50$. It is determined by the contribution of gauge-boson fluctuations plus the minimum matter contributions of three chiral fermion generations in the 16 representation and a complex scalar in the 10 representation which accounts for the Higgs doublet. The number $\mathcal{N}$ reflects the contributions from all additional matter fields, in particular the scalars needed for spontaneous breaking of the GUT-symmetry. We are interested in the regime $\mathcal{N}>\mathcal{N}_c$ where the modification of $\beta_\alpha$ by graviton fluctuations is needed in order to make the model viable.

\begin{figure}[!t]
\includegraphics[width=\linewidth, clip=true, trim=0.8cm 3.5cm 0cm 0cm]{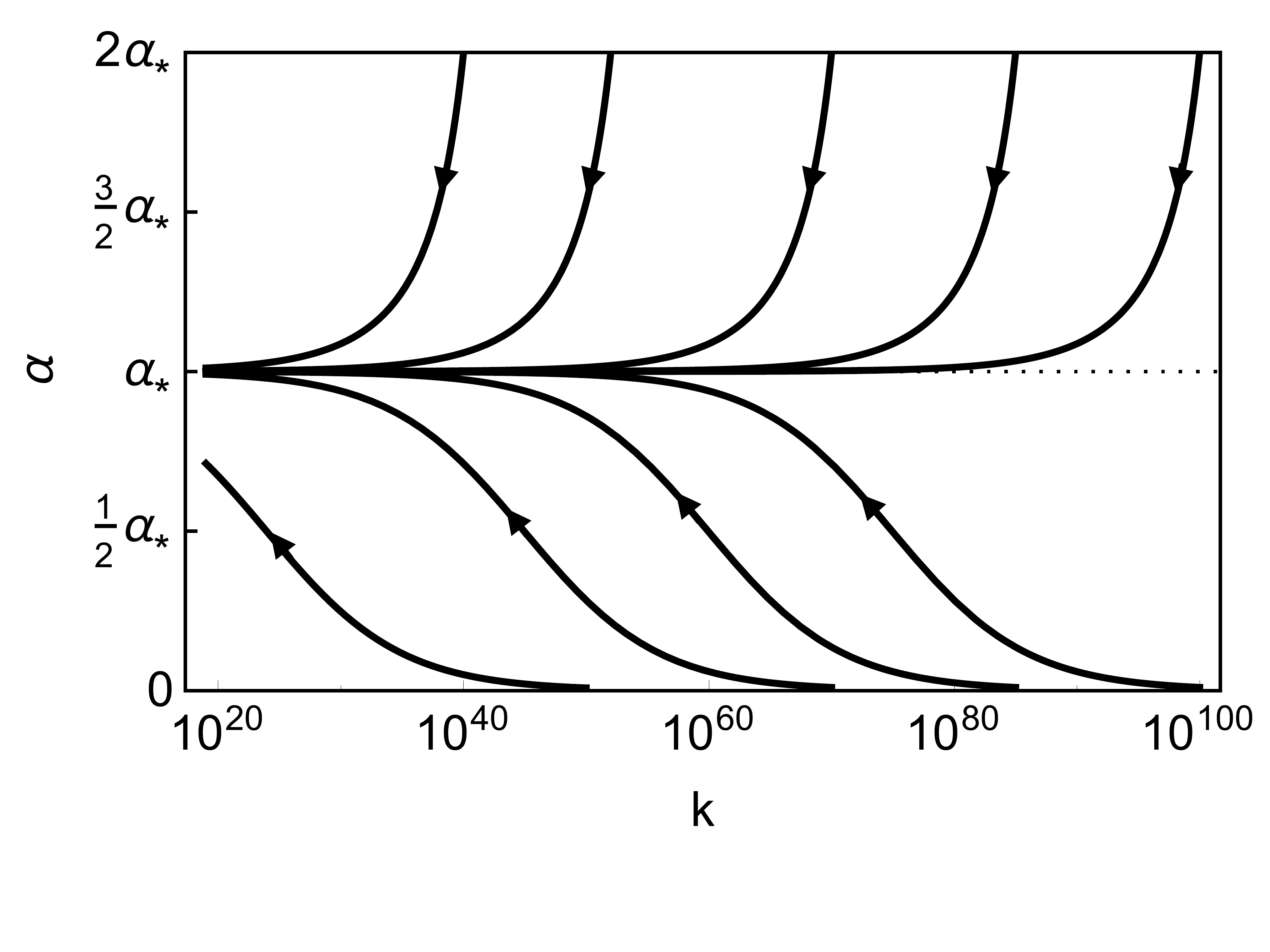}
\caption{\label{fig:illustration_ranges}
Flow of $\alpha$ as a function of the cutoff scale $k$. Finite values of $0<\alpha_1<\alpha_\text{UV}<\alpha_2$ at some ultraviolet scale $k_\text{UV}$ are mapped to an IR interval of allowed values for $k<k_\text{UV}$. This interval shrinks to the fixed-point value $\alpha_{\ast}$ for $k_{\rm UV}/k \rightarrow \infty$.
}
\end{figure}

\begin{figure}[!t]
\includegraphics[width=\linewidth, clip=true, trim=11.3cm 8cm 2cm 5cm]{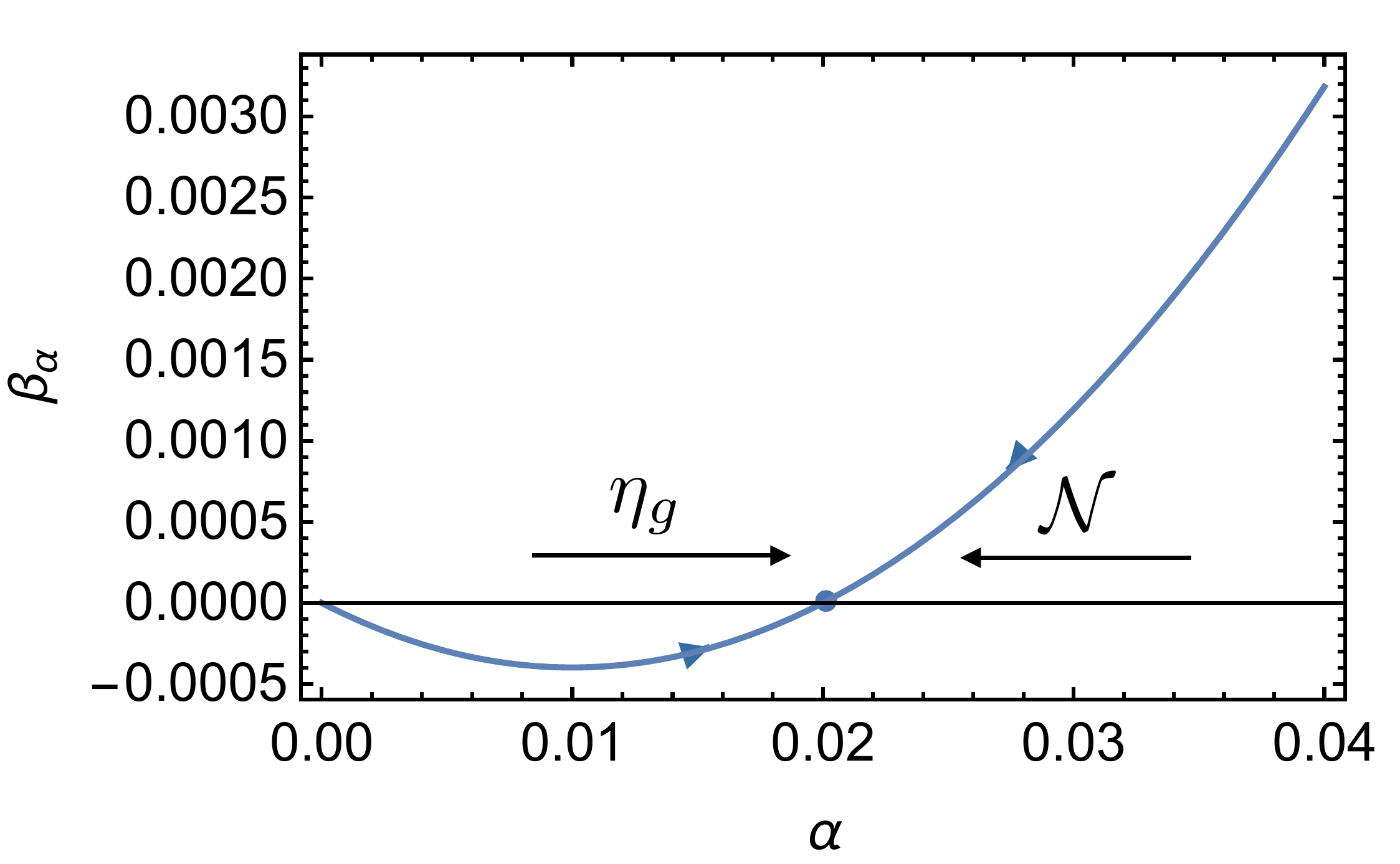}
\caption{\label{fig:illustration_beta} Flow generator $\beta_{\alpha}$ as a function of $\alpha$, evaluated for the fiducial value in eq.~\ref{eq:alphafpfid} and $\mathcal{N}=\mathcal{N}_c+50$. The arrows on the curve indicate the flow towards the IR. We also display the effect of strengthening gravity fluctuations, parameterized by $\eta_g$, and of increasing matter fluctuations, parameterized by $\mathcal{N}$.}
\end{figure}

The mechanism behind a prediction for the value of the gauge coupling is quite simple, and arises from a competition of quantum fluctuations of gravity and quantum fluctuations of matter fields and gauge bosons. 
At small values of the gauge coupling, the gravity contribution dominates. This contribution renders the gauge coupling asymptotically free 
\cite{Daum:2009dn,Folkerts:2011jz,Harst:2011zx,Christiansen:2017gtg,Eichhorn:2017lry,Christiansen:2017cxa}. 
In this regime the gauge coupling increases as the momentum scale is lowered, cf.~initial values $\alpha(k)<\alpha_{\ast}$ in Fig.~\ref{fig:illustration_ranges}.
In contrast, at large values of the gauge coupling, the matter 
contributions dominate and draw the gauge coupling towards smaller values, cf.~initial values $\alpha(k)>\alpha_{\ast}$ in Fig.~\ref{fig:illustration_ranges}.  
The antiscreening effect of gravity and gauge boson fluctuations and the screening effect of matter fluctuations cancel for a critical value $\alpha_{\ast}$ of the gauge coupling. This fixed point translates into a unique low-energy value of $\alpha$. \\
More precisely, the graviton fluctuations induce an anomalous dimension $\eta_g$ for the flow of $\alpha$, according to 
\be
\beta_{\alpha}= \eta_{g}\, \alpha+\left(\mathcal{N}-\mathcal{N}_c \right)\frac{\alpha^2}{4\pi}\;.\label{eq:betaalpha}
\ee
As illustrated in Fig.~\ref{fig:illustration_beta}, the system is dominated by the interplay of two fixed points: For $\eta_g<0$ the free fixed point at $\alpha=0$ is IR repulsive under the impact of gravity. Simultaneously, the presence of gravity induces an interacting fixed point at
\be
\alpha_{\ast} = -\frac{4\pi\,\eta_g}{\mathcal{N}- \mathcal{N}_c}.\label{eq:alphafp}
\ee
It is IR attractive, if
the gravity-induced anomalous dimension is 
negative
\be
\eta_g = -c_g \frac{k^2}{M^2(k)},
\ee
with  $M(k)$ the running Planck mass. The coefficient $c_g$ depends on further gravitational couplings, such as
the cosmological constant.
Explicit studies all indicate $c_g \geq  0$ \cite{Daum:2009dn,Folkerts:2011jz,Harst:2011zx,Christiansen:2017gtg,Eichhorn:2017lry,Christiansen:2017cxa}. In particular, the presumably dominant transverse traceless tensor (graviton) fluctuations yield a manifestly positive contribution \cite{Eichhorn:2017lry}. Uncertainties on the precise value remain due to different matter content and calculation schemes. Values quoted range between $\eta_g \approx -1.6$ \cite{Harst:2011zx} and $\eta_g \approx -0.1$ \cite{Eichhorn:2017lry}. 
An important part of this difference arises from the dependence of the fixed-point value 
of $k^2/M^2(k)$  on the number and nature of matter degrees of freedom. The value obtained from \cite{Harst:2011zx} refers to an Abelian gauge-gravity system, while \cite{Eichhorn:2017lry} takes into account the effect of minimally coupled standard model fields.\\
At the fixed point associated  with asymptotically safe quantum gravity one has $M_{\ast}(k)\sim k$ such that $\eta_g$  indeed takes a constant value $\eta_{g\, \ast}$  as expected for fixed-point behavior.
The value $k^2/M_{\ast}^2(k) = 8\pi G_{\ast}$ corresponds to the fixed-point value of the dimensionless Newton coupling $G$. The precise value of $G_{\ast}$ depends on the matter content \cite{Narain:2009fy,Dona:2013qba,Meibohm:2015twa,Dona:2015tnf, Eichhorn:2016vvy,Biemans:2017zca,Christiansen:2017cxa}, and uncertainties due to the computational scheme remain as well. 
A reasonable range estimate for our GUT setting may be $\eta_{g\ast}= -\frac{1}{4\pi}$, but we will leave this quantity undetermined in the following.\\
In the limit of fixed $\eta_g$, as realized at transplanckian scales, the flow eq.~\eqref{eq:betaalpha} has the solution
\be
\alpha(k)= \frac{ \alpha_{\ast}}{1+ \left(\frac{k_{\rm UV}}{k}\right)^{\eta_g}\left(\frac{\alpha_{\ast}}{\alpha_{\rm UV}} -1\right)},\label{eq:alpharun}
\ee
where we start at some scale $k_{\rm UV}$ with $\alpha(k_{\rm UV})= \alpha_{\rm UV}$. As $k_{\rm UV}$ increases, any interval 
$\alpha_1 \less \alpha_{\rm UV} \less \alpha_2$, with finite $\alpha_2,\alpha_1>0$, 
is mapped to smaller and smaller intervals at $k=M$. This is demonstrated in Fig.~\ref{fig:illustration_ranges}. For $k_{\rm UV} \rightarrow \infty$, this interval shrinks to the fixed point $\alpha(M)= \alpha_{\ast}$.  
In more detail, an upper bound for $\alpha(k)$ follows for $\alpha_{\rm UV} \rightarrow \infty$. In the limit $k_{\rm UV}/k \rightarrow \infty$ it coincides with $\alpha_{\ast}$. For $0< \alpha_{\rm UV}<\alpha_{\ast}$ one observes a crossover behavior between the two fixed points at 0 and $\alpha_{\ast}$. Since the asymptotically free fixed point at $\alpha=0$ is unstable towards the IR, any $\alpha_{\rm UV}>0$ reaches $\alpha_{\ast}$ at $k= M$ for $k_{\rm UV}/k\rightarrow \infty$. Quantum gravity is defined by a nonperturbative fixed point. If one chooses the one with $\alpha_{\ast}\neq 0$, the prediction $\alpha(M)=\alpha_{\ast}$ follows.

Below the Planck scale, gravity becomes sub-dominant very quickly, and the RG flow towards the IR is determined by quantum fluctuations of the fields in the GUT model, driving the gauge coupling away from the fixed-point value. For any given GUT model with a specific sequence of spontaneous symmetry breaking, the value $\alpha(M)$  required to reach the observed values in the IR differs slightly, but typical values lie close to $\alpha(M_{\rm GUT}) \approx 1/40$.
This value can be achieved by a balance between the gravity effect and the matter contribution, encoded in the two quantities $\eta_{g}$ and $\mathcal{N}$. 
Let us assume that the additional matter fields contributing to $\mathcal{N}$ are scalars in various representation of the GUT symmetry, as needed for spontaneous symmetry breaking. With $N_R$ the number of $R$-representations one has that \cite{Slansky:1981yr}
\bea
\mathcal{N} &=& \frac{1}{3}\Bigl(N_{10} +4N_{16}+8 N_{45}+12 N_{54}+ 28 N_{120}+70 N_{126}\nonumber\\
&{}& \quad+68 N_{144}+56 N_{210} \Bigr),
\eea
where all $N_R$ but $N_{126}$ refer to the number of real representations. 
For a realistic value $\alpha_{\ast} \approx 1/40$ and a value $\mathcal{N}-\mathcal{N}_c \approx 40$, the gravity-induced anomalous dimension must be of order $\eta_g \sim -\frac{1}{4\pi}$. An increase of $\mathcal{N}$ may lead to an increase of $-\eta_g$ and we leave open the question if for too large $\mathcal{N}$ the fixed point in the gravity sector disappears \cite{Dona:2013qba,Meibohm:2015twa,Dona:2015tnf, Eichhorn:2016vvy,Biemans:2017zca,Christiansen:2017cxa}.

For the case of SU(5), one finds $\mathcal{N}_c =85/3$ (including three chiral fermion generations in the $10$ and $\bar{5}$ and a complex scalar $5$ for the SM-Higgs) and
\bea
\mathcal{N} &=& \frac{1}{3}\Bigl(N_{5} +3N_{10} +7N_{15} +5 N_{24} +28 N_{35} +22 N_{40}\nonumber\\
&{}& \quad +24 N_{45} + 35 N_{50}+49 N_{70} \Bigr),
\eea
where all $N_R$ but $N_{24}$ refer to complex representations.

\subsection{Constraints for model-building}

For the above mechanism to work, the GUT model has to be in a regime where $\beta_{\alpha}$ is positive for large $\alpha$. This requires a sufficient number of matter fields to be added to the model beyond those representations containing the three generations of standard model fermions and the standard model Higgs.
Hence, there is a \emph{lower} bound on the number of matter fields in various representations of the GUT symmetry group,
\bea
\mathcal{N}_{\rm lower} &=& \mathcal{N}_c.
\eea 
The fixed-point value $\alpha_{\ast}$ is the maximal possible value for $\alpha (M)$ for a renormalizable QFT that can be extended to $k_{\rm UV} \rightarrow \infty$, cf.~eq.~\ref{eq:alpharun} and Fig.~\ref{fig:illustration_ranges}. Realistic phenomenology requires that $\alpha_{\ast}$ is not much smaller than $1/40$. Writing eq.~\eqref{eq:alphafp} in the form
\be
\alpha_{\ast} =\frac{ \eta_{g}}{\eta_{g\, \rm fiducial}}
\frac{1}{\mathcal{N}-\mathcal{N}_{c}}, \quad \eta_{g\, \rm fiducial}=-\frac{1}{4\pi},
\label{eq:alphafpfid}
\ee
we infer the upper bound
\bea
\mathcal{N}_{\rm upper}\left(\eta_g\right) &=&\frac{1}{\alpha_{\rm ph}(M)}\frac{ \eta_g}{\eta_{g\, \rm fiducial}}+\mathcal{N}_c,\label{eq:Nupper}
\eea
If the fixed point with $\alpha_{\ast}\neq 0$ is chosen for defining the ultraviolet completion of the quantum gravity gauge model, the value $\mathcal{N}_{\rm upper}$ is precisely the value for which a realistic gauge coupling is obtained.  
Here, $\alpha_{\rm ph}(M)$ is related to the observed gauge couplings by the renormalization flow between $M$ and low energy scales. The precise flow depends on the GUT-symmetry breaking mechanism and the content of particles with mass between $M$ and $M_{\rm GUT}$. For a large number of such particles $\alpha_{\rm ph}(M)$ is typically somewhat larger than 1/40, since the gauge coupling decreases between $M$ and $M_{\rm GUT}$.

For an example of complex scalars coupling to quarks and leptons, $N_{10}=6$, $N_{120}=4$, $N_{126}=1$ and real scalars $N_{54}=1$, $N_{45}=3$, $N_{210}=1$, where multiplicities may be motivated by a generation symmetry, one obtains $\mathcal{N}-\mathcal{N}_c =130/3$, leading to a realistic $\alpha_{\ast}$ for $\eta_g$ near $\eta_{g\, \rm fiducial}$.

\subsection{Discussion}
For a suitable matter content the interplay of grand unified gauge theories with quantum gravity renders the fine-structure constant predictable. While a precise value for this prediction has to wait for a quantitative settlement of the gravity-induced anomalous dimension $\eta_g$, several important conclusions can already be drawn at the present stage. The perhaps most important one is that an interacting fixed point in the flow of gravitational and gauge couplings establishes ``gravi-gauge theories'' as candidates for a consistent renormalizable QFT for all known interactions. Second, other couplings beyond the gauge coupling may become predictable. The basic mechanism is very simple, see, e.g., \cite{Eichhorn:2017egq}. Some couplings that correspond to relevant or marginal couplings for the flow at energies below $M$ (the usual renormalizable couplings) may actually be irrelevant couplings for the flow at energies above $M$, corresponding to irrelevant directions in the vicinity of the non-perturbative fixed point. Examples discussed so far are the quartic coupling of the Higgs doublet, whose very small fixed-point value translates to a Higgs mass around 126 GeV with a few GeV uncertainty \cite{Shaposhnikov:2009pv}, or the deviation of the mass term of the Higgs doublet from the critical value of the phase transition that may explain the hierarchy between the electroweak scale and the Planck scale by the resurgence mechanism \cite{Hamada:2017rvn}.

 An interacting fixed point underlying a quantitative determination of standard model parameters has been proposed for the Abelian hypercharge coupling in \cite{Harst:2011zx} and found to be in the vicinity of the standard-model value in \cite{Eichhorn:2017lry}.

Another example are predictable Yukawa couplings \cite{Eichhorn:2017ylw, Held:20175}.
The flow equation for the Yukawa coupling $y$ of the third generation (say between a scalar 10-plet and fermions in SO(10)) has a structure similar to eq.~\eqref{eq:betaalpha} for transplanckian scales $k$, 
\be
k\partial_k\, y^2 = \eta_y \, y^2 + a_y\, y^4\label{eq:betah},
\ee
with $a_y>0$. Both gauge boson and gravity fluctuations contribute to the anomalous dimension
\be
\eta_y = -b_y\, g^2 - c_y \frac{k^2}{M^2(k)}\label{eq:etah}.
\ee
If $\eta_y$ remains negative in the presence of the gravity contribution $c_y$ \cite{Eichhorn:2016esv,Oda:2015sma,Eichhorn:2017ylw,Eichhorn:2017eht}, one finds an interacting fixed point similar to eq.~\eqref{eq:alphafp}
\be
y_{\ast}^2 = \frac{-\eta_y}{a_y}\label{eq:hfp}.
\ee
At the fixed point one predicts the ratio
\be
\frac{y_{\ast}^2}{g_{\ast}^2} = \frac{1}{a_y}\left(b_y + \frac{4\pi\, c_{y}\, k^2}{\alpha_{\ast}\,M^2(k)} \right).\label{eq:fpratio}
\ee
For $k$ much smaller than $M$, the ratio $y^2/g^2$ starts deviating from the fixed-point value in eq.~\eqref{eq:fpratio} since the gravity fluctuations effectively decouple in eq.~\eqref{eq:etah}. The ratio $y^2/g^2$ is then attracted towards a partial fixed point \cite{Pendleton:1980as,Hill:1980sq,Wetterich:1981ir,Wetterich:1987az}, first to the one of the GUT-model, and after spontaneous GUT-symmetry breaking to the one of the standard model. As compared to the standard model with $g_{\ast}=0$, the additional negative contribution to $\eta_y$ from $g_{\ast}^2 \neq 0$ in GUT models favors the occurrence of this fixed point. 

The Yukawa coupling determined by the fixed point \eqref{eq:fpratio} typically determines the mass of the top quark. There are several possibilities for the quarks and leptons with smaller mass. If their Yukawa couplings correspond to the vicinity of the free fixed point $y_{\ast}=0$ of eq.~\eqref{eq:betah}, they cannot be predicted. Alternatively, a second Yukawa coupling, e.g., relevant for the bottom quark, may be determined by the fixed-point behavior in a more complex setting \cite{Held:20175}. Finally, the GUT-symmetry may be supplemented by a generation symmetry whose spontaneous breaking  can induce a small parameter $\lambda$. Mixings between doublets (with respect to the weak SU(2) subgroup) carrying different generation charges are suppressed by different powers of $\lambda$. They can be responsible for the hierarchical pattern of fermion masses and mixings \cite{Wetterich:1985fu,Bijnens:1987ff}. The analogous Frogatt-Nielsen mechanism \cite{Froggatt:1978nt} would require additional super-heavy fermion generations.

The way to a complete, realistic gravi-gauge theory still seems long. Important partial results, as in the present letter, should encourage an increased effort towards a more precise quantitative settlement of the flow contributions from gravity fluctuations.\\

\emph{Acknowledgments} \\
We thank J.~M.~Pawlowski for fruitful discussions.
A.~Eichhorn and A.~Held are supported by the Emmy-Noether-program of the DFG under grant no Ei-1037/1. A.~Held is also supported by a scholarship of the Studienstiftung des deutschen Volkes. A.~Eichhorn acknowledges support by an Emmy-Noether visiting fellowship of the Perimeter Institute for Theoretical Physics. C.~Wetterich is supported by the DFG collaborative research center SFB 1225 (ISOQUANT).

\end{document}